\documentclass[12pt,english,floatfix,superscriptaddress,aps,prd,preprint]{revtex4}
\usepackage{float}
\usepackage{array}
\usepackage{graphicx}
\usepackage{amsmath,amsthm,amsfonts,amssymb}
\usepackage[english]{babel}
\usepackage{color}
\usepackage[dvips]{epsfig}
\usepackage[dvips]{graphicx}
\usepackage{units}
\usepackage{gensymb}
\usepackage{xcolor}

\usepackage{hyperref}
\hypersetup{
    colorlinks,
    citecolor=blue,
    filecolor=green,
    linkcolor=purple,
    urlcolor=blue,
}

\begin{document}

\title{UHI in Fortaleza and trends on screen-level air temperature and precipitation}
\author{A. de A. Coelho}
\email{afranio@fisica.ufc.br}
\affiliation{Universidade Federal do Cear\'a (UFC), Departamento de F\'isica,\\ Campus do Pici, Fortaleza - CE, C.P. 6030, 60455-760 - Brazil.}
\date{\today}
\begin{abstract}
There is a consensus that Urban Heat Island phenomenon - UHI occurs in every large city. This effect is characterized by higher air temperatures in cities than in the neighboring countryside at night. However, to date, there has been no systematic study on the Fortaleza case, the Brazil's 5th largest city. By the comparison between screen-level air temperature measured by two automatic weather stations, one located in the city and the other in the neighboring region, this work shows the occurrence of the UHI in Fortaleza, even during the rainy season. In an attempt to find some effect of the UHI on precipitation (and vice-versa), historical series of air temperature and precipitation were analyzed from 1966 to 2020. Besides the considerable increase in air temperature over the years, a slight downward trend was observed in UHI and even more in precipitation between the hours of 15:00 and 21:00 (local time). However, one believes that these trends may be related to climate change at large scale rather than an urban scale.
\end{abstract}
\maketitle
\section{Introduction}
The fact that cities show higher temperatures than near rural landscape was first observed by Luke Howard research on the climate of London in the nineteenth century~\cite{Mills_2008}. This trend could be observed in all town and cities and emerge from differentiation on radiative fluxes and turbulent exchanges. Urban Heat Island -- UHI effect is characterized when closed isotherms separate the city from the general temperature field, producing a pattern like the contours of an ``island'' around the ``sea'' of the cooler surrounding countryside~\cite{Oke_1995}. This phenomenon is one of the clearest examples of unaware man-made climate change~\cite{Oke_2017}. Even though UHI is a human microclimatic change, this phenomena is highly depended on synoptic conditions in such a way that cloudy and windy weather attenuate it but clear and calm conditions reinforce it~\cite{Landsberg_1981}. UHI arises due to larger cities heat capacity than rural landscape. Indeed, paved surfaces and buildings walls retain much heat than green and clear surfaces during daytime. This absorbed heat is slowly released to air environment after sunset mainly due to the damming of the ground-emmited thermal radiation by tall buildings. Therefore, cities reach higher temperatures in the early evening hours than the neighboring rural zones.

Despite beneficial urban heat island effect in cold weathers as reducing need for heating, on the contrary UHI in hotter climates increases energy requirements for air-conditioning which retroactively enhances this effect. Moreover, UHI has harmful impacts on health, especially in the aggravation of diseases related to the respiratory system \cite{Hajat_2010}.

Several observations indicate that cities favor the formation of precipitation~\cite{Kratzer_1956};~\cite{Braham_1981};~\cite{Romanov_1999};~\cite{Ashley_2011};~\cite{Haberlie_2015}. Specifically, that precipitation tends to increase over and/or downwind of urban areas. One of the reasons is that the urban boundary layer - UBL is characterized by convergence and uplift, due to the effects of roughness and the urban heat island.

Considering that the authors are not aware of previous studies regarding the existence of the urban heat island phenomenon over the city of Fortaleza, this work intends to (1) demonstrate the occurrence of this phenomenon in the city and (2) extract temperature and precipitation trends over the years that may relate to the UHI evolution.
\section{Methodology}
The city of Fortaleza (3\degree 43\selectlanguage{english}' S, 38\degree 32\selectlanguage{english}' W), capital of the State of Ceara, is the fifth largest in Brazil. Located in the Northeast of Brazil, it has 314,930 km\textsuperscript{2} of total area, 2,643,247 inhabitants estimated in 2018, in addition to the highest demographic density among the Brazil's capitals, with 8,390.76 inhabitants per km\textsuperscript{2}. It comprises 34 km of the Atlantic coastline. Its average altitude is 16 m and has annual averages of temperature and precipitation around 27~\degree C and 1,700 mm, respectively. Figure 1 shows a location map of Fortaleza in Ceara state and in Brazil.\selectlanguage{english}
\begin{figure}[h!]
\begin{center}
\includegraphics[width=0.42\columnwidth]{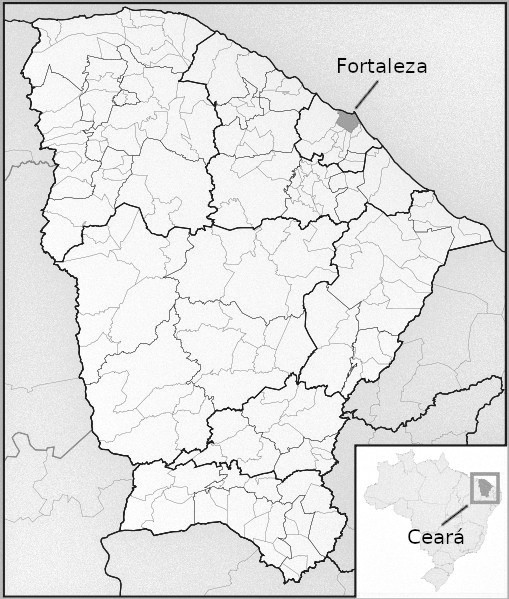}
\caption{\selectlanguage{ngerman}{Location map of the city of Fortaleza in Ceará State, Brazil.
{\label{Fig1}}%
}}
\end{center}
\end{figure}\selectlanguage{ngerman}

Intending to find the occurrence of the UHI phenomenon in Fortaleza, the present study measured hourly differences of thermal microscale between two paired automatic weather stations - AWS: one located at Fortaleza and the other on the country, at São Gonçalo do Amarante, a rural zone 47 km far to the northwest of Fortaleza, as shown by Fig. 2. The paired station approach is useful in the absence of an AWS network within and around a single urban area as was the case here~\cite{Morris_2001}. One can see from this satellite image the contrast between the green area of the neighboring rural region and the built-up area of the city of Fortaleza, in the lower right corner of the image below.\selectlanguage{english}
\begin{figure}[h!]
\begin{center}
\includegraphics[width=0.70\columnwidth]{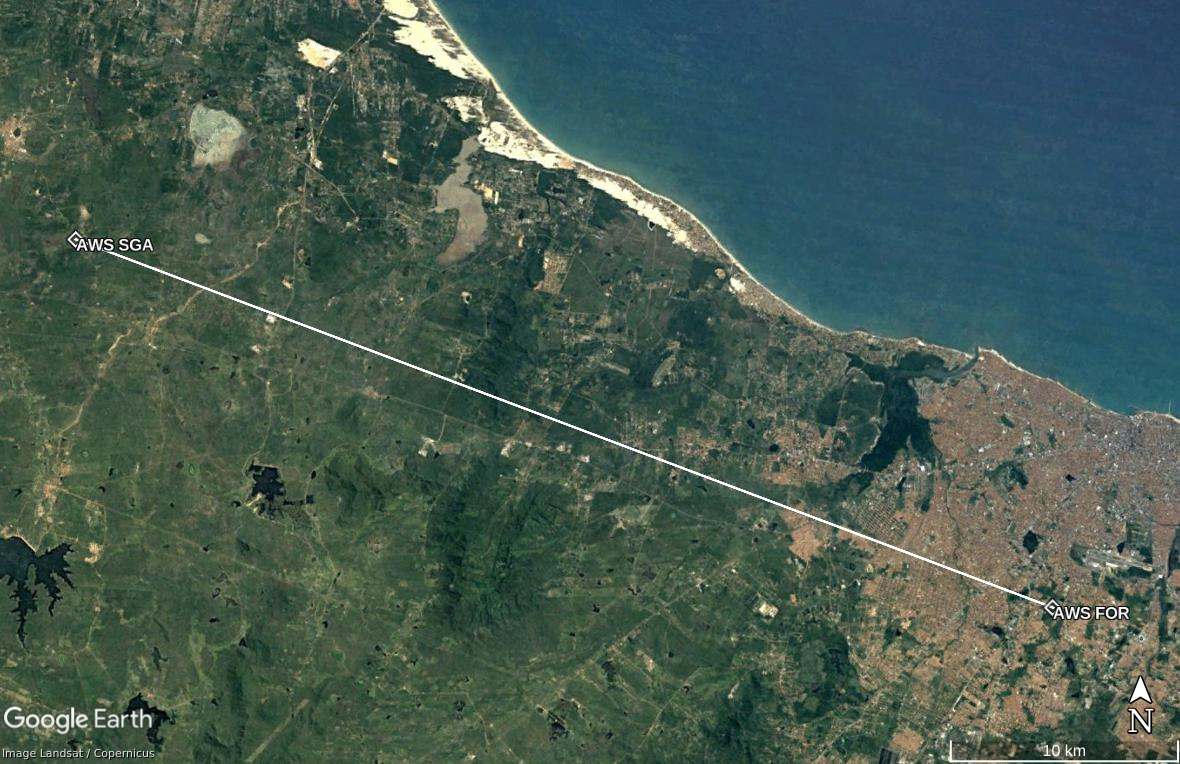}
\caption{\selectlanguage{ngerman}{Satellite view from 2003 of the region where the automatic stations in Fortaleza (AWS FOR) and São Gonçalo do Amarante (AWS SGA) were located. Aerial photograph courtesy of Google Earth.
{\label{Fig2}}%
}}
\end{center}
\end{figure}\selectlanguage{ngerman}

The thermometers were installed at two identical Campbell Scientific AWS. Figure 3 shows the AWS located in Fortaleza (3\degree 47' 42'' S, 38\degree 33' 30'' W; referred to hereafter as AWS FOR) on an university campus of the State University of Ceará - UECE. Unfortunately, there is no photo from the AWS installed in São Gonçalo do Amarante (3\degree 39' 83'' S, 38\degree 56' 32'' W; referred to hereafter as AWS SGA). Both of the AWS were located in open environments with vegetated surface features similar to those found in natural landscapes. These AWS belong to and are maintained by the Ceará meteorology and water resources foundation (Fundação Cearense de Meteorologia e Recursos Hídricos - FUNCEME). All Funceme´s AWS follow the same site selection criterion and are equipped with the same type of sensors. The temperature sensor is a Vaisala HMP45A probe with ±0.2 °C accuracy at +20 °C. These sensors were calibrated in the factory and there were no post-purchase calibration procedures.\selectlanguage{english}
\begin{figure}[h!]
\begin{center}
\includegraphics[width=0.70\columnwidth]{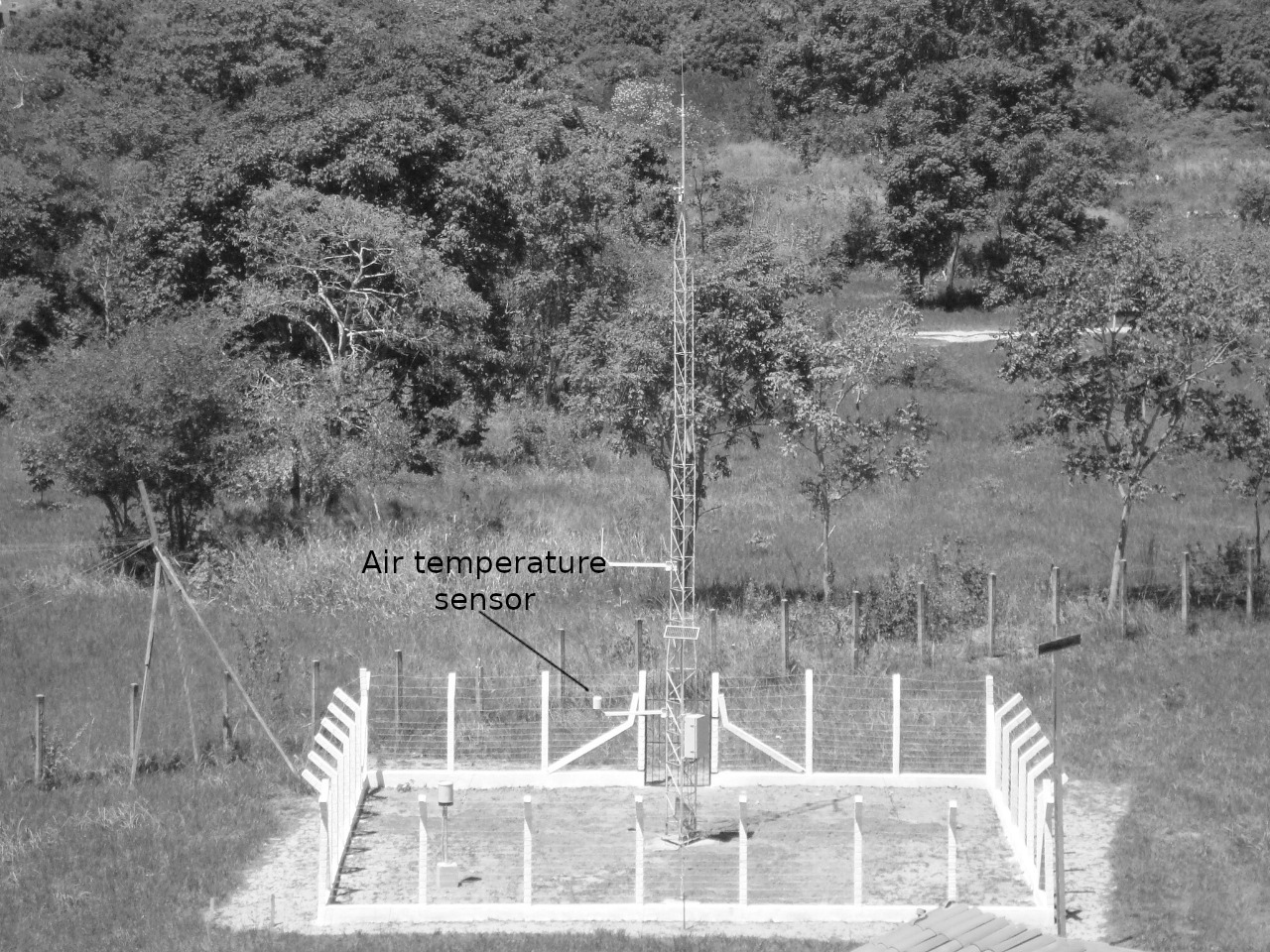}
\caption{{Campbell Scientific weather station located in Fortaleza. Photography courtesy of FUNCEME.
{\label{Fig3}}%
}}
\end{center}
\end{figure}
Figure 4 shows satellite visible imagery from 2006 provided by Google Earth of these locations where the stations are in the center of two concentric circles: the inner one with a radius of 500 m and the outer one with a radius of 1.5 km.\selectlanguage{english}
\begin{figure}[h!]
\begin{center}
\includegraphics[width=0.84\columnwidth]{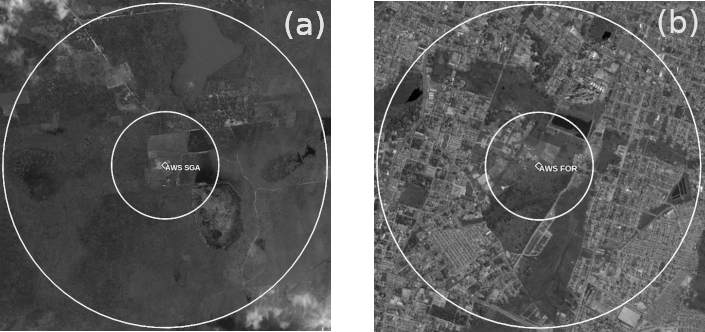}
\caption{\selectlanguage{ngerman}{Satellite images of the automatic weather station settlements: (a) in São Gonçalo do Amarante - AWS SGA (3\degree 39' 83'' S, 38\degree 56' 32'' W); (b) in Fortaleza - AWS FOR (3\degree 47' 42'' S, 38\degree 33' 30'' W). The
weather stations are in the center of two concentric circles: the inner one with a radius of 500 m and the outer one with a radius of 1.5 km. Aerial photographs courtesy of Google Earth.
{\label{Fig4}}%
}}
\end{center}
\end{figure}\selectlanguage{ngerman}

In order to allow inter-site comparisons for urban heat island studies, and to reinforce an objective protocol for reporting on UHI intensity, the climate of the places where the two stations were installed was classified according to the Local Climate Zone - LCZ classification system proposed by Stewart and Oke~\cite{Stewart_2012}. The LCZ  is defined by the authors as a ``region of uniform surface cover, structure, material, and human activity that span hundreds of meters to several kilometres in horizontal scale''. This definition can be applied to the area where the AWS SGA is located. In effect, the
surface structure and land cover remain quite homogeneous for three kilometres, as can be seen from the aerial photograph in Figure 4a. On the contrary, the urban environment of Fortaleza shows a high level of heterogeneity where LCZ classes are visibly discrete, such that the boundaries separating the classes can be delineated, as can be seen from the the aerial photograph in Figure 4b.

Since it was not possible to collect appropriate site metadata to quantify the surface properties of the source areas, only these aerial photographs were used to classify these field sites into LCZs. Thus, the so-called rural area was classified in this study as a LCZ B
(\emph{Scattered trees}) characterized by ``lightly wooded landscape of deciduous and/or evergreen trees. Land cover mostly pervious (low plants)'' \cite{Stewart_2012}. On the other side, a LCZ B\textsubscript{9} sublclass was created for the the so-called urban area, where  LCZ B is the higher parent class and LCZ 9 (\emph{Sparsely built}) is the lower parent class from the standard set which is typified by ``sparse arrangement of small or medium-sized buildings in a natural setting. Abundance of pervious land cover (low plants, scattered trees)''~\cite{Stewart_2012}.

According to Oke~\cite{Oke_1988}, in a context of separation between two zones with different surface patterns (the~\emph{fetch effect}) where air passes from one surface-type to a new and climatically different surface, it must adjust to a new set of boundary conditions. When the air flows from a zone with a smooth and bare surface to a fully moist low vegetation cover, the complete adjustment of the boundary layer to the properties of the vegetated surface occurs at a rate of 100 to 300 m for every 1 metre increase in the vertical. Knowing that the measurements of air temperature and humidity in the two AWSs were made at a height of 2 meters, the boundary layer would be fully adjusted to this height to an upwind fecth at about 200 to 600 m. Considering that the border areas are not smooth and bare where the AWSs were installed, it can be assumed that this adjustment distance may occur at a shorter distance. Therefore, each LCZ has a radius of 500 m so that the adjusted part of its fully adjusted boundary layer is within the zone and does not overlap with surrounding LCZs of different structure or coverage. The adoption of this criterion can be critical
for the situation of the AWS FOR as the neighboring surfaces can have significant climatic differences. In fact, from Fig. 4b one can see the predominance of dense settlement patterns beyond the 500 m radius circle that refer to a LCZ classification of type 3 (LCZ 3: compact low-rise) with ``dense mix of low-rise buildings. Few or no trees. Land cover mostly paved. Stone, brick, tile, and concrete construction materials''~\cite{Stewart_2012}.

As mentioned above, one of the goals of the present work is to find temperature and precipitation trends in Fortaleza and analyzing them from the perspective of the influence of the UHI. In this sense, the comparison between the hourly temperature measured in the two AWS was made only for the rainy season, which comprises the months of February to May, in the years 2004 to 2006. Therefore, the methodology applied did not only consider ideal weather situations ~\cite{Oke_2017}, especially on cloudless days. Indeed, the rainy season during this period is mainly caused by the Intertropical Convergence Zone - ITCZ southward displacement from the tropical Atlantic Ocean which is characterized by light winds.

With regard to the physiographic setting of this UHI treatment, the two thermometers were located in a flat orography on the coast of the South Atlantic (see Figure 2). The urban heat island featured in this study is a typical canopy layer UHI -- UHI\textsubscript{UCL}~\cite{Oke_1995} since the air temperature differences was measured from two fixed platforms: one in the Urban Canopy Layer -- UCL (the air layer beneath roof level) in the city; the other in a corresponding height (2 m) over the non-urban landscape.
\section{UHI Results}
Figure 5 shows the average hourly (local time) air temperature observed in AWS FOR and AWS SGA during the rainy season from 2004 to 2006. The data includes all the 121 days of 2004 (February to May), 101 days of 2005 and only 13 days of 2006. The lower number of days in 2005 and especially in 2006 is due to the malfunction of one of the stations on certain days. Therefore, only the days when it was possible to perform the complete pairing of reliable measurements over the 24 hours were selected. The average heat island magnitude \(\overline{\Delta{T}}_{U-R}\) represents the difference between these average temperatures and is calculated simply by:
\begin{equation}
\overline{\Delta{T}}_{U-R}=\overline{T}_U-\overline{T}_R,
\end{equation}
where $\overline{T}_U\) and \(\overline{T}_R$ are the average air temperature in the urban area (AWS FOR) and rural area (AWS SGA), respectively. 

One of the most striking aspects of the canopy layer heat island can be shown in Fig. 5, namely its predominance at night. This because urban areas cool more slowly at night than neighboring rural areas~\cite{Oke_2017}. A wilcoxon signed rank pairwise test indicated that $\overline{T}_U\) and \(\overline{T}_R$ were significantly different to each other at the 95\% confidence level for most of the day except at 7:00 am, as indicated by the square around bullet point in Fig. 5.

Although Fortaleza presents a typical scenario of evolution of the UHI throughout the day, some aspects are worth noting. Indeed, $\overline{T}_U$ catches up to $\overline{T}_R$ later than midday around sunset. As these are measurements during the rainy season, the soil in the rural region must have been very wet. Precipitation will then increase the thermal admittance of the more permeable rural surface, thus reducing nocturnal cooling, leading to lower magnitudes of UHI~\cite{Runnalls_2000}. In addition, due to frequent low cloud cover in this period, it is very likely that the maximum magnitude of $\overline{\Delta{T}}_{U-R}$ is underestimated~\cite{Morris_2001}. Indeed, the maximum UHI measured during this period reached up to 5~°C which occurred on a night (at 3:00 am on March 11, 2006) with no rain and light wind. UHI reaches its peak around 23:00 and slowly
decreases perhaps because of the known increase in precipitation after this time, which keeps the rural soil wet and also promotes the imprisonment in the city of the latent heat released by the condensation of rain droplets. Then, it declines quickly after sunrise and this drop ends after 10:00. After dawn, rural warming quickly erodes the nocturnal UHI and the daily sequence repeats. Sunrise and sunset over these months happened around 6:00 and 18:00, respectively.\selectlanguage{english}
\begin{figure}[h!]
\begin{center}
\includegraphics[width=0.70\columnwidth]{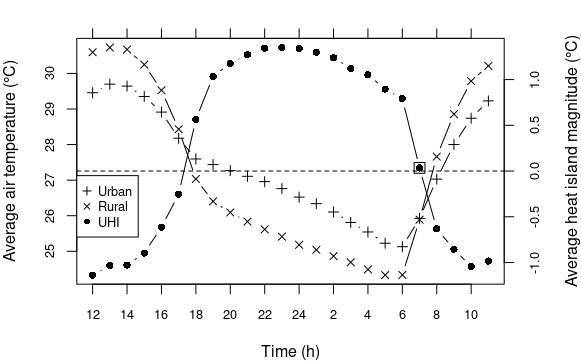}
\caption{{Mean air temperature in AWS FOR (+) and AWS SGA (x) per hour (local
time) throughout the day and the respective UHI magnitude
(\(\bullet\)); a bullet enclosed by square indicates differences
non-statistically significant different from zero at the 95\% confidence
level.
{\label{Fig5}}%
}}
\end{center}
\end{figure}
Despite its particularities, it can be concluded from Figure 5 that the urban heat island effect occurred in the LCZ where the AWS FOR is installed, even during the rainy season from 2004 to 2006. Unfortunately, there were insufficient data for each of the two AWS to pair screen-level temperature for other years.

In order to confirm this result, these same air temperature data (the same period) from the AWS SGA were compared with temperature measurements made at a manual weather station (MWS) inside the campus of the Federal University of Cear\selectlanguage{ngerman}á - UFC (3\degree 44' 20'' S, 38\degree 35' 37'' W; hereafter referred to as MWS UFC). Unfortunately, no other weather automatic station is available in the urban area of Fortaleza to obtain a broader representation of the UHI profile within the city. The MWS UFC has provided data on temperature, precipitation, wind speed, among others, since 1966. The data is regularly collected at three times of the day: at 9:00, 15:00 and 21:00 (local time). Precipitation data are accumulated rain that fell, if any, between these times of day. Figure 6 shows satellite visible satellite imagery from 2003 provided by Google Earth of the location of the AWS FOR (Fig. 6a) and the MWS UFC (Fig. 6b) in Fortaleza. As can be seen when comparing the locations of the two weather stations in Fortaleza, the MWS UFC is located in an area with more built-up spaces, including a large urban power distribution system on the west side. This mix of different surface structure and cover hinders a clear and unambiguous assignment for the LCZ, especially without the availability of metadata. Based only on aerial imagery from that time, It seems that more than half the area of the 500 m radius circle around the AWS UFC (see Figure 7b) can be classified as an LCZ 9 (\emph{sparsely built}) which consists of a ``sparse arrangement of small or medium-sized buildings in a natural setting and abundance of pervious land cover (low plants, scattered trees)''~\cite{Stewart_2012}. The rest of the area can be typified as a LCZ 3 (\emph{compact low-rise}). Thus, the entire area is subclassified as a LCZ 9\textsubscript{3}.\selectlanguage{english}
\begin{figure}[h!]
\begin{center}
\includegraphics[width=0.84\columnwidth]{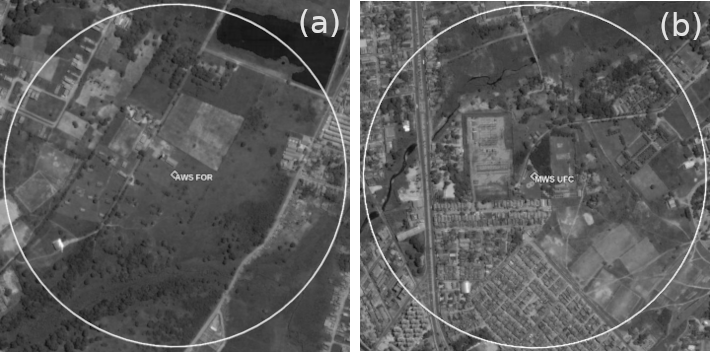}
\caption{{Satellite image of the location of the two weather stations in
Fortaleza: (a) the automatic weather station - AWS FOR
(3\degree 47' 42'' S, 38\degree 33' 30'' W); (b)
the manual weather station - MWS UFC (3\degree 44' 20'' S,
38\degree 35' 37'' W). Both weather stations are located in
the center of a circle with a radius of 500 m. Aerial photographs
courtesy of Google Earth.
{\label{Fig6}}%
}}
\end{center}
\end{figure}
Average values of temperature and differences are listed in Table 1 between MWS UFC and the two automatic weather stations: AWS SGA and AWS FOR. Pearson's coefficient correlation between the data is also listed. The average temperature differences (statistically significant at 95\% confidence) indicate that the UHI phenomenon also occurs at 21:00 in the MWS UFC climate zone, while it is absent at 9:00 and 15:00 in the day, as noted in AWS FOR (see Figure 5). As the MWS UFC and AWS FOR temperature time series show a strong correlation (correlation coeficiente greater than 0.8), the average UHI magnitude is slightly lower than in the case of the comparison between AWS FOR and AWS SGA (see Table 1). This is an unexpected outcome as the climate zone where the manual station is located has more built-up area (see Fig. 6b). In fact, the average air temperature in the MWS UFC is a little lower at 21:00, but higher at 9:00 and 15:00. This indicates that the AWS FOR local zone behaves as a kind of  `heat islet' compared to the MWS UFC location. This may relate to an inaccurate survey of measurements or the source area for the sensor is not fully contained within the internal boundary layer (IBL) of the corresponding LCZ. Therefore, one can affirm that, as in all large cities, the UHI phenomenon was identified in Fortaleza in at least two of its different climate zones. Although this evidence applies to measurements made nearly twenty years ago, one can assume that this effect has remained over the years, mainly due to the continued growth of the city. In this regard, an attempt was made to extract the trend of air temperature evolution in Fortaleza in order to verify any aspect of the UHI trend over the years, as described below.\selectlanguage{english}
\begin{table}[h!]
\centering
\normalsize
\resizebox{\textwidth}{!}{%
\begin{tabular}{ c c c c c c c c c c }
   \hline
   & 09:00 &  &  & 15:00 &  &  & 21:00 &  &  \\
   \hline
   & MWS UFC & AWS SGA & AWS FOR & MWS UFC & AWS SGA & AWS FOR & MWS UFC & AWS SGA & AWS FOR \\
   \hline
$\overline{T}$ & 28.3 & 28.9 & 28 & 29.7 & 30.2 & 29.4 & 26.9 & 25.8 & 27.1 \\
$\Delta \overline{T}$ &  & -0.6 & 0.3 &  & -0.5 & 0.3 &  & 1.1 & -0.2 \\
Correlation &  & 0.68 & 0.85 &  & 0.79 & 0.87 &  & 0.65 & 0.84 \\
\hline
\end{tabular}}
\caption{\selectlanguage{ngerman}{The average air temperature (in \degree C) observed in the manual weather station in Fortaleza (MWS UFC) and in the automatic weather stations in São Gonçalo do Amarante (AWS SGA) and Fortaleza (AWS FOR), and the difference. Pearson's coefficient correlation between the data is also listed.
{\label{Tab1}}%
}}
\end{table}\selectlanguage{ngerman}
\section{Air temperature and precipitation trends}
The MWS UFC annual average climate data from 1966 to 2020 (i.e., 55 observations per year) was used to determine intrinsic trend in screen-level air temperature in Fortaleza. Given the explicit definition of trend as ``an intrinsically fitted monotonic function or a function in which there can be at most one extremum within a given data span''~\cite{Wu_2007}, the trends were extracted by applying the Empirical Mode Decomposition - EMD method~\cite{Huang_1998}. It decomposes time series into Intrinsic Mode Functions - IMF which are oscillations on various time scales where the number of local minima and maxima differs at most by one and have a mean value of zero. The residue of this process is a monotonic function with only one set of extremes, which is the trend as defined above. The EMD method is more suitable than the traditional ones for the analysis of non-linear and non-stationary time series and has been proved to be extremely powerful in extracting simple components from a given signal~\cite{Stallone_2020}. In summary, the EMD is an adaptive method that will decompose a data set \(x\left(t\right)\) in terms of IMF \(x_n\left(t\right)\) and a residue \(r\left(t\right)\) so that the
signal can be represented by:
\begin{equation}
x\left(t\right)=\sum_n^{ }x_n\left(t\right)+r\left(t\right).
\end{equation}

Unfortunately, it was not possible to carry out a comparative analysis of the trend extracted in Fortaleza with another obtained in a non-urban area, such as São Gonçalo do Amarante. This was because on the one hand, the AWS SGA only operated for a few years, which makes it impossible to extract any statistically significant trend from any time series obtained by this station. On the other hand, conventional meteorological stations located near or in the region of São Gonçalo do Amarante have meteorological data collected in the early morning, at 7:00. As well know, the UHI is a nocturnal phenomenon. So, conventional observational data may be of no value to test hypotheses related to the UHI phenomenon ~\cite{Lowry_1998}.

In this context, trend comparisons were performed only with data recorded by the MWS UFC. Data obtained at 9:00 (Morning - M) and 15:00 (Afternoon - A) were considered not influenced by the UHI, as opposed to data from 21:00 (Nocturnal - N) when the UHI is well established (see Figure 5 and Table 1). As it is data collected from the same weather station, one can assume it to be equally affected by trends at larger (greater than urban) scales~\cite{Lowry_1977}. Therefore, trends in their difference are likely to be significant and may be associated with effects caused by the UHI.

Figure 7 shows mean annual air temperature values based on long-term measurements over 55 years during the rainy season (from February to May) at the MWS UFC. Symbols and dashes in red and green in Figure 7a refer to measurements taken at 09:00 and 15:00, respectively, while those in blue refer to measurements taken at 21:00. The difference between the mean annual values and the respective trend is shown in Figure 7b. Despite the strong correlation between these values, the temperature rise between the 70s and 80s seems to have been earlier and more intense in the evening than in the daytime. The greatest variation in demographic density (52\%) recorded in Fortaleza occurred during these two decades, when the city had a remarkable horizontal growth. Thereafter, the population and the city continued to grow, but at a slower rate. Since the 1990s, a verticalization reached residential sections east of the city centre ~\cite{Costa_2015}.\selectlanguage{english}
\begin{figure}[h!]
\begin{center}
\includegraphics[width=0.91\columnwidth]{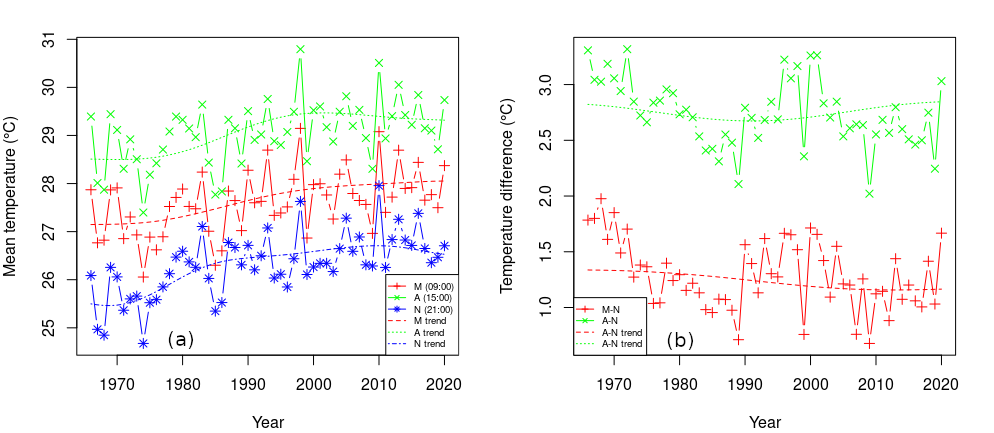}
\caption{{Long-term variation of (a) annual mean air temperature for MWS UFC in daytime (red and green) and at night (blue), and (b) their difference.
{\label{Fig7}}%
}}
\end{center}
\end{figure}
Nevertheless, the transformations in the urban order that took place over these years in Fortaleza seem to have not caused an accentuated trend of variation between daytime and night temperatures. By doing the difference between daytime and night temperature time series, any UHI trend could be revealed. In fact, no trend seems to exist compared to afternoon temperatures but a slight decreasing trend in UHI is identified in relation to morning temperatures (see Fig. 7b). This because the upward trending is more remarkable for morning than nighttime temperatures, as can be seen from Fig7a. Furthermore, morning temperatures are less correlated with nighttime than afternoon temperatures. In any case, it can be speculated that the UHI phenomenon in Fortaleza did not evolve far beyond modulations generated by larger atmospheric scales or global climate change.

The impact of climate change on the UHI is addressed by a systematic review conducted by Chapman et al. (2017) on scientific papers published between 2000 and 2016 based on a modelling approach. They found ``wide variation responses of the UHI to climate change that can be explained by changes in the factors that lead to the development of UHIs, such as weather, soil dryness and anthropogenic heat'' ~\cite{Chapman_2017}. Soil dryness decreases the UHI by reducing rural evaporative fraction and decreasing the urban--rural temperature difference. The cause of this was the prediction by Lemonsu et al. (2013) of a substantial decrease of the urban heat islands in Paris by reduced precipitation in the surrounding countryside that would dry out natural soils by severely limiting evaporation, which tends to strengthen the sensible heat flow~\cite{Lemonsu_2013}.

To analyze the effects of precipitation on UHI in Fortaleza, we sought to determine, in a first approach, the trend in the temporal evolution of rainfall over the years during the night period. Figure 8 shows annual precipitation values based on long-term measurements over 55 years during the rainy season (from February to May) for the MWS UFC at 21:00 (accumulated rainfall between 15:00 and 21:00, local time). It can be seen that there is a well-defined trend towards a decrease in accumulated rainfall between 15:00 and 21:00: about 36\% over this period. Unfortunately, it is not possible to make a rural-urban
comparison of the evolution of the precipitation due to absence of historical data and that precipitation measurements are collections of rain fallen in the last 24 hours, as mentioned above. Thus far, it can be assumed that the slight decrease of the UHI in AWS FOR location, as shown above (see Fig. 7a), may be caused by soil drying as predicted by Lemonsu et al. (2013). No trend was found for the period between 21:00 and 9:00.\selectlanguage{english}
\begin{figure}[h!]
\begin{center}
\includegraphics[width=0.70\columnwidth]{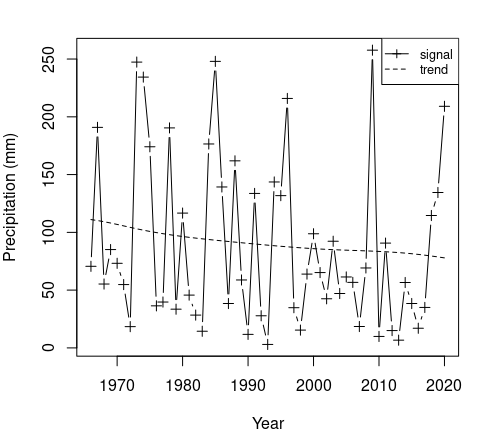}
\caption{{Long-term variation of annual accumulated rainfall between 15:00 and 21:00 for MWS UFC and the trend.
{\label{Fig8}}
}}
\end{center}
\end{figure}
\section{Conclusions}
In spite of limitations in obtaining hourly temperature data in the surrounding countryside of Fortaleza, it was possible to identify the UHI phenomenon in this city during the rainy season, from February to May. The UHI appears just after sunset, around 18:00, and vanishes at sunrise, around 6:00. In order to analyze the influence of the UHI on precipitation and vice-versa, historical temperature and precipitation records (from 1966 to 2020) were analyzed in search of trends. Despite the existence of an increasing trend in the average annual temperature in the morning, in the afternoon and in the night in Fortaleza, especially in the 80's and 90's, a comparative analysis between daytime and nighttime (when the UHI phenomenon occurs) temperatures did not show any marked UHI trend in Fortaleza, except for a slight decrease in the difference between the temperatures in the morning and in the night. This decrease may be associated with the increase in temperatures in the morning as well as the possible increase in soil dryness at the weather station site. In fact, a decreasing trend was observed over these years in the accumulated rainfall between 15:00 and 21:00. However, it is believed that this trend is not associated with the UHI, given that this effect only occurs after 18:00. Furthermore, between 21:00 and 09:00, a period more affected by the UHI, no apparent trend was found. Thus, it is concluded that any trend found is most likely due to changes in weather at large scale rather the urban one.

Future studies of this type should be undertaken to give insight into precipitation effects on UHI (and vice-versa) notably with the support of remote sensing systems (e.g. weather radar and satellite microwave).
\section*{Acknowledgments}
\hspace{0.5cm} The author thanks the Ceará Meteorology and Water Resources Foundation - FUNCEME, in particular M. Sc. Wagner Luiz Barbosa Melciades, the Programa de Bolsa de Iniciação Acadêmica - BIA of the Federal University of Ceará - UFC, in particular Mr. Douglas da Silva Teixeira, and the colleagues who helped in the review of this work, in particular Dr. Jean Iaquinta.


\end{document}